\documentclass[journal]{IEEEtran}

\ifCLASSINFOpdf
\else
\fi

\usepackage{setspace}%????
\usepackage{bbm}
\usepackage[cmex10]{amsmath}
\usepackage{amssymb}
\usepackage{cite}
\usepackage{graphicx}
\usepackage{array,color}
\usepackage{amsmath}
\allowdisplaybreaks
\usepackage{stfloats}
\usepackage{graphicx}
\usepackage{epstopdf}
\usepackage{subfigure}
\usepackage{tabularx}
\usepackage{epsfig,epsf,color,balance,cite}
\usepackage{algorithmic}
\usepackage{algorithm}
\usepackage{url}
\usepackage{bm}
\usepackage{multirow}

\usepackage{amsthm}
\ifodd 1
\usepackage{soul}
\usepackage{color}
\setstcolor{red}

\newcommand{\del}[1]{\st{#1}} %deleting the text

\newcommand{\com}[1]{\textbf{\color{red} (COMMENT: #1)}} %comment of the text
\newcommand{\response}[1]{\textbf{\color{green} (RESPONSE: #1)}} %response to comment
\else

\newcommand{\del}[1]{}

\newcommand{\com}[1]{}
\newcommand{\comg}[1]{}
\newcommand{\response}[1]{}
\fi
\title {\vspace{-0.10cm}Uplink Channel Estimation for Double-IRS Assisted Multi-User MIMO}

\author{ \vspace{-0.15cm}\normalsize
	Beixiong Zheng, Changsheng You, and Rui Zhang \\
	\normalsize Department of Electrical and Computer Engineering, National University of Singapore, Singapore\\
	Email: \{elezbe, eleyouc, elezhang\}@nus.edu.sg
	
\vspace{-1.0cm}
}

\begin{document}
%\markboth{Submitted to IEEE Transations on XXX}{SKM: My IEEE article}
\maketitle
%\vspace{-1cm}
\begin{abstract}
%To achieve higher-order beamforming gains of the cooperative IRS, accurate channel state information
%(CSI) is indispensable but practically challenging to acquire, 
%Although the double-IRS cooperative shows to achieve higher-order beamforming gain than the conventional single-IRS system, the 
%The higher-order beamforming gain of the double-IRS system generally comes at the expense of extra channel coefficients to be estimated, which 
%and thus calls for high-efficient channel estimation schemes. 
%In the double-intelligent reflecting surface (IRS) assisted system with fully passive IRSs and the co-existence of the double- and single-reflection links, the channel estimation problem is practically challenging and thus calls for innovative solutions.
To achieve the more promising passive beamforming gains in the double-intelligent reflecting surface (IRS) assisted system over the conventional single-IRS system, channel estimation is practically indispensable but also a more challenging problem to tackle, due to the presence of not only the single- but also double-reflection links that are intricately coupled.
In this paper, we propose a new and efficient channel estimation scheme for the double-IRS assisted uplink multiple-input multiple-output (MIMO) communication system to resolve the cascaded channel state information (CSI) of both its single- and double-reflection links. 
First, for the single-user case, the higher-dimensional double-reflection channel is efficiently estimated at the multi-antenna base station (BS) with low training overhead by exploiting the fact that its cascaded channel coefficients are scaled versions of those of a lower-dimensional  single-reflection channel.
Then, the proposed channel estimation scheme is extended to the multi-user case, where given an arbitrary user's cascaded channel estimated as in the single-user case, the other users' cascaded channels are scaled versions of it and thus can be estimated with reduced training overhead.
Simulation results verify the effectiveness of the proposed channel estimation scheme as compared to the benchmark scheme.
\end{abstract}
%\vspace{-0.1cm}
%\begin{IEEEkeywords}
%	Intelligent reflecting surface (IRS), distributed IRSs, channel estimation, training reflection design, multi-user multiple-input multiple-output (MIMO), mean squared error (MSE).
%\end{IEEEkeywords}
\IEEEpeerreviewmaketitle

\vspace{-0.5cm}
\section{Introduction}
%\IEEEPARstart{I}{ntelligent}
Intelligent reflecting surface (IRS) is an innovative solution to the realization of smart
and reconfigurable environment for wireless communications \cite{wu2020intelligent,qingqing2019towards,Renzo2019Smart}. Specifically, IRS consists of a large number of passive reflecting 
elements with ultra-low power consumption, each of which is capable of controlling the phase shift and/or amplitude of the incident signal in a programmable manner so as to collaboratively reshape the wireless propagation channel in favor of signal transmission. Moreover, as being light weight and free of radio frequency (RF) chains, large-scale IRS can
be densely deployed in various wireless communication systems \cite{Zheng2020IRSNOMA,Pan2020Multicell,yang2019intelligent} with a low and scalable energy consumption and implementation cost.

\begin{figure}[!t]
		\centering
		\includegraphics[width=2.5in]{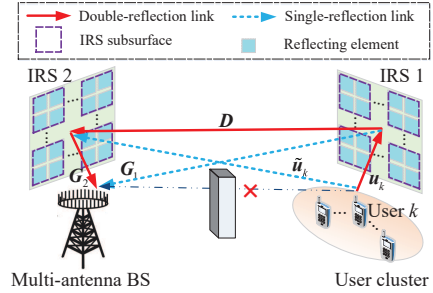}
		\setlength{\abovecaptionskip}{-6pt}
		\caption{A double-IRS assisted multi-user MIMO communication system in the uplink.}
		\label{system}
		\vspace{-0.7cm}
	\end{figure}

Prior works on IRS mainly considered the wireless communication systems assisted by one or more distributed IRSs,
each independently serving its nearby users without taking into account the inter-IRS signal reflection, which, however, fails to capture the cooperative beamforming gains between IRSs to further improve the system performance. Only recently,
the cooperative beamforming gains over the inter-IRS channel has been explored in the double-IRS assisted system \cite{Han2020Cooperative,Zheng2020DoubleIRS,you2020wireless}, which was shown to achieve a much higher-order passive beamforming gain than the conventional single-IRS system (i.e., ${\cal O}(M^4)$ versus ${\cal O}(M^2)$ with $M$ denoting the total number of reflecting elements in both systems). However, achieving such a more appealing passive beamforming gain requires more channel training overhead in practice, due to more channel coefficients to be estimated over the inter-IRS double-reflection link, in addition to the single-reflection links in the conventional single-IRS system.
Existing works on IRS channel estimation mainly focused on the channel state information (CSI) acquisition for single-reflection links only \cite{zheng2019intelligent,zheng2020intelligent,zheng2020fast,you2019progressive,wang2019channel,jensen2019optimal}, which, however, is inapplicable to the double-IRS assisted system with the co-existence of single- and double-reflection links as illustrated in Fig.~\ref{system}.
In \cite{Han2020Cooperative}, the authors assumed that the two IRSs are equipped with receive RF chains to enable
	the sensing capability for estimating their channels with the base station (BS)/user, separately. Nonetheless, even with receive RF chains integrated to IRSs, the channel estimation for the inter-IRS (i.e., IRS~1$\rightarrow$IRS~2) link is still practically difficult. 
	%their nature of passive reflection without the capability of transmitting any pilot signals between them.
	In contrast, the double-IRS channel estimation with fully passive IRSs was investigated in \cite{you2020wireless}, but without the single-reflection links considered and for the single-user case only. 
	%which, however, suffers from much higher-order training overhead.
	%the fully passive IRSs without any receiving/transmitting capability, which is more practical with much lower power consumption and implementation cost yet more challenging for channel estimation.

	To overcome the above issues, we propose in this paper a new and efficient channel estimation scheme for the double-IRS assisted multi-user multiple-input multiple-output (MIMO) system shown in Fig.~\ref{system}, 
	where the communications between a multi-antenna BS and a cluster of nearby users are assisted by two fully passive IRSs, which are deployed near the BS and the cluster of users, respectively.
%	To overcome the above issues, we consider in this paper a double-IRS cooperative system shown in Fig.~\ref{system}, 
%	where the communications between a multi-antenna BS and a cluster of nearby users are assisted by two fully passive IRSs. Under this setup, we propose a new and efficient channel estimation scheme for the single-user case to acquire all the cascaded channel coefficients.
	%where a key property (channel correlation relationship) that the single- and double-reflection links reflected by IRS~2 share the same (common) IRS~2$\rightarrow$BS channel is exploited.
	% for both the single- and double-reflection links.
	% the channel correlation property between them is exploited due to the common shared IRS~2$\rightarrow$BS channel
	%In particular, the key idea of the proposed channel estimation lies in 
	%In particular, the channel correlation property due to the common shared IRS~2$\rightarrow$BS channel is explored to achieve  
	%In particular, by exploiting a key property that the single- and double-reflection links reflected by IRS~2 share the same (common) IRS~2$\rightarrow$BS channel, the training overhead 
	%Under this setup, to avoid co-channel interference, we consider the decoupling channel estimation for estimating the single- and double-reflection links separately.
	% with the leverage of simple ON/OFF IRS reflection. 
	First, for the single-user case, the cascaded channels of the two single-reflection links, each corresponding to one of the two IRSs respectively, are successively estimated at the multi-antenna BS with the other IRS turned OFF. 
	%Then, by exploiting the channel correlation property due to the common shared IRS~2$\rightarrow$BS channel, we can estimate the double-reflection (i.e., user$\rightarrow$IRS~1$\rightarrow$IRS~2$\rightarrow$BS) link efficiently after taking the estimated (user$\rightarrow$IRS~2$\rightarrow$BS) channel as reference CSI, which thus achieve much less training overhead.
	%% and taking the estimated user$\rightarrow$IRS~2$\rightarrow$BS channel as the reference CSI, 
	Then, after canceling the signals over the two single-reflection channels estimated,
	the higher-dimensional double-reflection (i.e., user$\rightarrow$IRS~1$\rightarrow$IRS~2$\rightarrow$BS) channel is efficiently estimated at the BS
	%with substantially reduced training overhead 
	by exploiting the fact that its cascaded channel coefficients (through each subsurface of IRS~1) are the scaled versions of those of the single-reflection (i.e., user$\rightarrow$IRS~2$\rightarrow$BS) channel due to their commonly shared IRS~2$\rightarrow$BS link;
	as a result, only the lower-dimensional scaling factors need to be estimated for the double-reflection channel, which substantially reduces the training overhead.
	Next, the proposed channel estimation scheme is extended to the general multi-user case, where given an arbitrary user's cascaded channel estimated as in the single-user case, the other users' cascaded channels are scaled versions of it and thus can be estimated with reduced training overhead.
	%by exploiting the channel correlation relationship among different users as in \cite{wang2019channel}.
	It is shown by simulation that the proposed channel estimation scheme achieves lower training overhead and also improves channel estimation performance as compared to the existing scheme based on \cite{you2020wireless}.
	\vspace{-0.3cm}
\section{System Model and Problem Description}\label{sys}
%\subsection{Double-IRS Assisted Multi-User MIMO}
% composed of $K$ reflecting elements 
Consider a double-IRS assisted multi-user MIMO communication system shown in Fig.~\ref{system}, in which the communications between a cluster of $K$ single-antenna users and an $N$-antenna BS are assisted by two distributed IRSs (referred to as IRS~1 and IRS~2).
%with a large number (denoted by $N$) of  antennas.
%To minimize the path loss between the IRSs and their associated BS/users, we assume the practical deployment scenario where IRSs~1 and 2 are placed near the cluster of users and the BS, respectively.   
As in \cite{Zheng2020DoubleIRS}, the direct links between the users and the BS are assumed to be blocked due to obstacles (e.g., walls/corners in the indoor environment).\footnote{If the direct links are non-negligible, 
	the BS can estimate them by the conventional channel estimation method with
	orthogonal/sequential pilots sent by the users and the two IRSs both turned OFF.}
To overcome the blockage as well as minimize the path loss, IRSs~1 and 2 are placed near the cluster of users and the BS, respectively,
such that the $K$ users can be effectively served by the BS through the single- and double-reflection links created by them.
%, i.e., users$\rightarrow$IRS~1$\rightarrow$IRS~2$\rightarrow$BS.
Let $M$ denote the total number of passive subsurfaces for the two distributed IRSs, where IRSs~1 and 2 comprise $M_1$ and $M_2$ subsurfaces, respectively, with $M_1+M_2=M$.
%To reduce the training overhead as well as simplify passive beamforming design, we adopt the element-grouping strategy as in \cite{zheng2019intelligent,yang2019intelligent,zheng2020intelligent}, which groups adjacent IRS elements a subsurface
%As in \cite{zheng2019intelligent,yang2019intelligent,zheng2020intelligent}, we adopt the element-grouping strategy to group every $\nu$ adjacent IRS elements into a subsurface for sharing a common reflection coefficient, thus greatly reducing the training overhead and design complexity.
%Specifically, the total number of $M$ passive reflecting elements are divided into $M={\tilde M}/\nu$ subsurfaces, where IRSs~1 and 2 consist of $M_1={\tilde M}_1/\nu$ and $M_2={\tilde M}_2/\nu$ subsurfaces, respectively, with $M_1+M_2=M$.
By adopting the element-grouping strategy in \cite{yang2019intelligent,zheng2019intelligent}, each of these IRS subsurfaces is composed of an arbitrary number of adjacent reflecting elements that share a common phase shift for reducing the channel estimation and reflection design complexity.
% thus enjoying a high aperture gain but with significantly reduced cost for channel estimation and reflection optimization, which generally increases with the number of subsurfaces.
%including multiple adjacent IRS elements 
%Moreover, each distributed IRS is connected to a smart controller that adjusts its phase shifts and exchanges information with the BS via a separate reliable wireless link \cite{qingqing2019towards}. 
In this paper, we assume the quasi-static flat-fading channel model for all the channels during each channel coherence interval.

Let ${{\bm u}}_{k}\triangleq\left[u_{k,1},\ldots, u_{k,M_1}\right]^T\in {\mathbb{C}^{M_1\times 1}}$, ${\tilde{\bm u}}_{k}\triangleq\left[{\tilde u}_{k,1},\ldots, {\tilde u}_{k,M_2}\right]^T\in {\mathbb{C}^{M_2\times 1}}$, ${{\bm D}}\triangleq\left[{\bm d}_{1},\ldots, {\bm d}_{M_1}\right] \in {\mathbb{C}^{M_2\times M_1 }}$, ${{\bm G}}_1 \in {\mathbb{C}^{N\times M_1 }}$,
and ${{\bm G}}_2 \in {\mathbb{C}^{N\times M_2 }}$ 
denote the baseband equivalent channels in the uplink for the user~$k$$\rightarrow$IRS~1, user~$k$$\rightarrow$IRS~2, IRS~1$\rightarrow$IRS~2, IRS~1$\rightarrow$BS, and IRS~2$\rightarrow$BS links, respectively, with $k=1,\ldots,K$. 
%Let ${\bm \theta}_\mu\triangleq[{\theta_{\mu,1}},{\theta_{\mu,2}},\ldots,{\theta_{\mu,M_\mu}}]^T=[ \beta_{\mu,1}e^{j \phi_{\mu,1}},  \beta_{\mu,2}e^{j \phi_{\mu,2}},\ldots, \beta_{\mu,M_\mu} e^{j \phi_{\mu,M_\mu}}]^T$ with $\mu\in \{1,2\}$ denote the equivalent reflection coefficients of IRS $\mu$, where $\phi_{\mu,m} \in [0, 2\pi)$ represents the phase shift of subsurface $m$ at IRS $\mu$ and the reflection amplitudes of all subsurfaces are set to one (i.e., $\forall \beta_{\mu,m}=1$) or the maximum value during the data transmission for simplicity.
Let ${\bm \theta}_\mu\triangleq[{\theta_{\mu,1}},\ldots,{\theta_{\mu,M_\mu}}]^T=\left[ \beta_{\mu,1}e^{j \phi_{\mu,1}},  \ldots, \beta_{\mu,M_\mu} e^{j \phi_{\mu,M_\mu}}\right]^T$ denote the equivalent reflection coefficients of IRS $\mu$ with $\mu\in \{1,2\}$, where $\beta_{\mu,m} \in [0, 1]$ and $\phi_{\mu,m} \in [0, 2\pi)$ are
the reflection amplitude and phase shift of subsurface~$m$ at IRS~$\mu$, respectively.
%we consider the binary ON/OFF control for the reflection amplitude, i.e., $\beta_{\mu,m} \in \{0,1\}, \forall m=1,2,\ldots,M_\mu, \mu\in \{1,2\}$ in the two IRSs for the channel training design.
%, where $|{\theta_{\mu,m}}|=1, \forall m=1,\ldots,M_\mu$, $\mu\in \{1,2\}$.
%\footnote{The reflection amplitudes of all subsurfaces are set to one or the maximum value to maximize the signal reflection power.}
%To maximize the signal power reflected by the IRS and reduce the hardware cost, we set $\beta_m=1, \forall  m\in {\cal M} $ and only consider the phase-shift design of the IRS.
%Thus, the effective channel from user~$k$ to the BS by
%combining the double-reflection link
% (i.e., user~$k$$\rightarrow$IRS~1$\rightarrow$IRS~2$\rightarrow$BS channel)\footnote{Although there exists another double-reflection link over the user~$k$$\rightarrow$IRS~2$\rightarrow$IRS~1$\rightarrow$BS channel, it suffers from much higher path loss due to the much longer propagation distance (see Fig.~\ref{system}) and thus is ignored in this paper.} 
% and the two single-reflection links (i.e., user~$k$$\rightarrow$IRS~2$\rightarrow$BS and user~$k$$\rightarrow$IRS~1$\rightarrow$BS channels) is given by
% and the cascaded BS$\rightarrow$IRS$\rightarrow$user channel, denoted by ${\bm h} \in {\mathbb{C}^{L\times 1 }}$, 
Thus, the effective channel from user~$k$ to the BS is the superposition of the double-reflection link and the two single-reflection links (see Fig.~\ref{system}), which is given by
\vspace{-0.2cm}
\begin{align}
	{\bm h}_k=&{{\bm G}}_2 {\bm \Phi}_2 {{\bm D}} {\bm \Phi}_1 {{\bm u}}_{k} 
	+{{\bm G}}_2 {\bm \Phi}_2 {\tilde{\bm u}}_{k} 
	+ {{\bm G}}_1 {\bm \Phi}_1 {{\bm u}}_{k} \label{superposed0}
	%\\
	%=&{{\bm G}}_2 {\bm \Phi}_2  {{\bm D}} {\bm \Phi}_1 {{\bm u}}_{k} 
	%+ {\tilde{\bm R}}_{k}{\bm \theta}_2
	%+ {{\bm R}}_{k}{\bm \theta}_1,  \quad k=1,\ldots,K\label{superposed}
	%\\
	%=&{{\bm G}}_2 {\bm \Phi}_2 
	%\left({{\bm D}} \text{diag} \left( {{\bm u}}_{k}  \right){\bm \theta}_1+ {\tilde{\bm u}}_{k}\right).\label{superposed2}
\end{align}
where ${\bm \Phi}_\mu=\text{diag} \left( {\bm \theta}_\mu \right)$ denotes the diagonal reflection matrix of IRS $\mu$ with $\mu\in \{1,2\}$.
%, and the cascaded user~$k$$\rightarrow$IRS~1$\rightarrow$BS channel (without phase shifts of IRS 1). 
Since we consider the fully passive IRSs without any receiving/transmitting capability, it is infeasible to acquire the CSI between the two IRSs as well as that with the BS/users separately. Nonetheless, it was shown in \cite{Zheng2020DoubleIRS} that the cascaded CSI (to be specified below) is sufficient for the cooperative reflection/passive beamforming design of the two IRSs to maximize the data transmission rate without loss of optimality.
% (to be specified later)
%Furthermore, by turning 
%IRS~2 OFF (i.e., $\forall \beta_{2,m}=0$), CSI of the cascaded user~$k$$\rightarrow$IRS~1$\rightarrow$BS link (i.e., ${{\bm R}}_{k}$) and the direct user~$k$ $\rightarrow$BS (i.e., ${{\bm u}}_{d,k}$) can be acquired by applying the existing single-IRS channel estimation methods in \cite{zheng2019intelligent,yang2019intelligent,wang2019channel} with the minimum training overhead of $M_1+2K-1$ time slots for the total number of $K$ users. Thus, in the rest of this paper we mainly focus on the more challenging channel estimation for the remaining part after performing the cancellation of ${{\bm R}}_{k}{\bm \theta}_1+{{\bm u}}_{d,k}$ in \eqref{superposed}, i.e., 
%\begin{align}
%{\bm h}_k&={{\bm G}}_2 {\bm \Phi}_2 \left( {{\bm D}} {\bm \Phi}_1 {{\bm u}}_{k} 
%+ {\tilde{\bm u}}_{k} \right) \notag\\
%&={{\bm G}}_2 {\bm \Phi}_2 
%\left({{\bm D}} \text{diag} \left( {{\bm u}}_{k}  \right){\bm \theta}_1+ {\tilde{\bm u}}_{k}\right).
%\end{align}
As such, let ${{\bm R}}_{k}={{\bm G}}_1 \text{diag} \left( {{\bm u}}_{k} \right)\in {\mathbb{C}^{N\times M_1 }}$ (${\tilde{\bm R}}_{k}={{\bm G}}_2   \text{diag} \left( {\tilde{\bm u}}_{k} \right)\in {\mathbb{C}^{N\times M_2 }}$) denote the cascaded user~$k$$\rightarrow$IRS~1 (IRS~2)$\rightarrow$BS channel (without taking the effect of IRS reflection yet), and 
${\tilde{\bm D}}_k\triangleq\left[{\tilde{\bm d}}_{k,1},\ldots, {\tilde{\bm d}}_{k,M_1}\right]={{\bm D}} \text{diag} \left( {{\bm u}}_{k} \right)\in {\mathbb{C}^{M_2\times M_1 }}$ denote the cascaded user~$k$$\rightarrow$IRS~1$\rightarrow$ IRS~2 channel (without taking the effect of IRS reflection yet) with ${\tilde{\bm d}}_{k,m}={\bm d}_{m} {{u}}_{k,m}, \forall m=1,\ldots,M_1$.
Then, the channel model in \eqref{superposed0} can be equivalently expressed as
%${\tilde{\bm \theta}}_1=\begin{bmatrix}1\\{\bm \theta}_1\end{bmatrix}$
\vspace{-0.2cm}
\begin{align}
	&{\bm h}_k
	={{\bm G}}_2 {\bm \Phi}_2  {\tilde{\bm D}}_k  {\bm \theta}_1
	+ {\tilde{\bm R}}_{k} {\bm \theta}_2
	+ {{\bm R}}_{k}{\bm \theta}_1  \notag\\
	=&{{\bm G}}_2 \left[{\bm \Phi}_2{\tilde{\bm d}}_{k,1},\ldots, {\bm \Phi}_2{\tilde{\bm d}}_{k,M_1} \right] {\bm \theta}_1+ {\tilde{\bm R}}_{k} {\bm \theta}_2
	+ {{\bm R}}_{k}{\bm \theta}_1 \notag\\
	=&{{\bm G}}_2 \hspace{-0.1cm}\left[\hspace{-0.05cm} \text{diag} \hspace{-0.1cm}\left({\tilde{\bm d}}_{k,1}\right)\hspace{-0.1cm}{\bm \theta}_2,\ldots, \text{diag} \hspace{-0.1cm}\left({\tilde{\bm d}}_{k,M_1}\right)\hspace{-0.1cm}{\bm \theta}_2 \right] {{\bm \theta}}_1\hspace{-0.1cm}+\hspace{-0.1cm} {\tilde{\bm R}}_{k}{\bm \theta}_2
	\hspace{-0.1cm}+ \hspace{-0.1cm}{{\bm R}}_{k}{\bm \theta}_1 \notag\\
	=&\sum\limits_{m=1}^{M_1}  \underbrace{{{\bm G}}_2 ~\text{diag} \left({\tilde{\bm d}}_{k,m}\right)}_{{{\bm Q}}_{k,m}} {\bm \theta}_2 {\theta}_{1,m}
	+ {\tilde{\bm R}}_{k}{\bm \theta}_2
	+ {{\bm R}}_{k}{\bm \theta}_1
	\label{superposed3}
\end{align}
where ${{\bm Q}}_{k,m}\in {\mathbb{C}^{N\times M_2 }}$ denotes the cascaded user~$k$$\rightarrow$IRS~1$\rightarrow$IRS~2$\rightarrow$BS channel associated with subsurface~$m$ at IRS~1, $\forall m=1,\ldots,M_1$ for the double-reflection link.
According to
\eqref{superposed3}, it is sufficient to acquire the cascaded CSI of $\left\{ {{\bm Q}}_{k,m} \right\}_{m=1}^{M_1}$, ${\tilde{\bm R}}_{k}$, and ${{\bm R}}_{k}$ for jointly designing the passive beamforming $\left\{{\bm \theta}_1, {\bm \theta}_2\right\}$ for the uplink data transmission in the double-IRS assisted system \cite{Zheng2020DoubleIRS}.
However, in practice,
the total number of channel coefficients in $\left\{ {{\bm Q}}_{k,m} \right\}_{m=1}^{M_1}$, ${\tilde{\bm R}}_{k}$, and ${{\bm R}}_{k}$ is prohibitively large, which consists of two parts:
\begin{itemize}
	\item The number of channel coefficients (equal to $K \times NM_1M_2$) for the high-dimensional double-reflection link (i.e., $\left\{ {{\bm Q}}_{k,m} \right\}_{m=1}^{M_1}$), which are newly introduced due to the double IRSs.
	\item  The number of channel coefficients (equal to $K \times N(M_1+M_2)$) for the two single-reflection links (i.e., ${{\bm R}}_{k}$ and ${\tilde{\bm R}}_{k}$), which exist in the conventional single-IRS assisted system (with either IRS 1 or IRS 2 present).
\end{itemize}
%, which is generally much larger than that for the conventional single IRS deployment (equal to $K\times (NM_2+NM_2)$). 
As can be seen, 
the number of channel coefficients for the double-reflection link is of higher-order than that for 
the two single-reflection links due to the fact that $M_1M_2\gg M_1+M_2$ in practice, which makes the channel estimation problem more challenging for the double-IRS assisted system, as compared to the single-IRS counterpart.
Note that given a channel coherence interval, such a considerably larger number of channel coefficients may require  significantly more training overhead that renders much less or even no time for data transmission, thus resulting in reduced achievable rate of the double-IRS assisted system (despite the higher passive beamforming gain over the double-reflection link assuming perfect CSI as shown in \cite{Zheng2020DoubleIRS}). 

To tackle the above challenge, we propose a new and efficient channel estimation scheme for the double-IRS assisted system to achieve minimum training overhead.
In particular, we exploit the channel relationship between the double-reflection link $\left\{ {{\bm Q}}_{k,m} \right\}_{m=1}^{M_1}$ and the single-reflection link ${\tilde{\bm R}}_{k}$, due to the same (common) IRS~2$\rightarrow$BS channel (i.e, ${{\bm G}}_2$) shared by them to reduce training overhead.
We first consider the single-user setup, i.e., $K=1$, to illustrate the main idea of the proposed channel estimation scheme for the double-reflection link in Section \ref{ON/OFF}, and then extend the results to the general multi-user case in Section \ref{MU_ON/OFF}.  
To reduce the hardware cost, we consider the binary ON/OFF control for the training reflection amplitudes of the two IRSs, i.e., $\beta_{\mu,m} \in \{0,1\}, \forall  m=1,\ldots,M_\mu,\mu\in \{1,2\} $ in our proposed channel estimation scheme.

%The BS then applies a linear receive beamforming vector ${\bm w}_k^H \in {\mathbb{C}^{N\times 1}}$ 
%to decode each $x_k$, i.e., 
%\begin{align}
%{\tilde y}_k=&{\bm w}_k^H\sum_{j=1}^{K} \left(\sum_{m=1}^{M_1} {{\bm Q}}_{j,m} {\bm \theta}_2 {{\theta}}_{1,m}+{\tilde{\bm R}}_{j} {\bm \theta}_2
%+ {{\bm R}}_{j}{\bm \theta}_1 \right) s_j + {\bm w}_k^H{\bm v}.
%\end{align}
%Then, the SINR for decoding $x_k$ is given by
%\begin{equation}\label{SINR}
%	\gamma_k=\frac{ \left|{\bm w}_k^H\left(\sum_{m=1}^{M_1} {{\bm Q}}_{k,m} {\bm \theta}_2 {{\theta}}_{1,m}+{\tilde{\bm R}}_{k} {\bm \theta}_2
%		+ {{\bm R}}_{k}{\bm \theta}_1\right)\right|^2}
%	{  \left|{\bm w}_k^H\sum_{j\neq k}\left(\sum_{m=1}^{M_1} {{\bm Q}}_{j,m} {\bm \theta}_2 {{\theta}}_{1,m}+{\tilde{\bm R}}_{j} {\bm \theta}_2
%		+ {{\bm R}}_{j}{\bm \theta}_1\right)\right|^2+\sigma^2 {\bm w}_k^H {\bm w}_k}
%\end{equation}
%Moreover, the achievable rate of user $k$ is
%\begin{align}\label{rate}
%R_k=\frac{T-\tau}{T}\log_2(1+\gamma_k),  \quad k=1,\ldots,K
%\end{align}
%where $\tau$ denotes the number of time slots for channel training.
\vspace{-0.2cm}
\section{Channel Estimation for Single-User Case}\label{ON/OFF}
In this section, we study the cascaded channel estimation for the single-user case with $K=1$. 
For notational convenience, the user index $k$ is omitted in this section.

%\subsection{Channel Estimation for Single User}

%With all the subsurfaces at IRS~2 turned OFF, the channel model in \eqref{superposed3} is reduced to the single IRS reflection channel, i.e., ${\bm h}={{\bm R}}{\bm \theta}_1$. As such, the existing single-IRS channel estimation methods in \cite{zheng2019intelligent,yang2019intelligent,wang2019channel} can be applied to acquired the CSI of ${{\bm R}}$
%with the minimum training overhead of $M_1$ time slots.
%We first turn OFF IRS~2 
%by turning 
%IRS~2 OFF (i.e., $\forall \beta_{2,m}=0$), CSI of the cascaded user$\rightarrow$IRS~1$\rightarrow$BS link (i.e., ${\tilde{\bm Q}}$) and the direct user~$k$ $\rightarrow$BS (i.e., ${{\bm u}}_{d,k}$) can be acquired by applying the existing single-IRS channel estimation methods in \cite{zheng2019intelligent,yang2019intelligent,wang2019channel} with the minimum training overhead of $M_1+2K-1$ time slots for the total number of $K$ users.
%In this subsection, we consider the channel estimation of an arbitrary selected user (say user $k$) with $N\ge M_2, M_1$ to illustrate the basic idea and draw some useful insights. 

According to \eqref{superposed3}, the cascaded user$\rightarrow$IRS~1$\rightarrow$IRS~2$\rightarrow$BS channel
through each subsurface $m$ at IRS~1 is given by
%\footnote{For national simplicity, we refer the cascaded user$\rightarrow$IRS~1$\rightarrow$IRS~2$\rightarrow$BS channel 
%through the virtual subsurface $0$ at IRS~1 to as the cascaded user$\rightarrow$IRS~2$\rightarrow$BS channel since this channel ${\tilde{\bm R}}$ also involves the common channel ${{\bm G}}_2$.}
\vspace{-0.2cm}
\begin{align}\label{cascaded_Q}
{{\bm Q}}_{m}={{\bm G}}_2 ~\text{diag} \left({\tilde{\bm d}}_{m}\right), \quad m=1,\ldots, M_1.
\end{align}
%where ${\tilde{\bm R}}\triangleq{\tilde{\bm R}}$ and ${\tilde{\bm u}}\triangleq{\tilde{\bm u}}$.
%Intuitively, it seems that there are $NM_2M_1$ channel coefficients in $\left\{ {{\bm Q}}_{m} \right\}_{m=1}^{M_1}$ to be estimated.
It is observed that all ${{\bm Q}}_{m}$'s in \eqref{cascaded_Q} share the same (common) IRS~2$\rightarrow$BS channel (i.e, ${{\bm G}}_2$) as ${\tilde{\bm R}}$.
As such,
if given the single-reflection channel ${\tilde{\bm R}}={{\bm G}}_2   \text{diag} \left( {\tilde{\bm u}} \right)$ as the reference CSI, we can re-express \eqref{cascaded_Q} as
\vspace{-0.2cm}
\begin{align}\label{cascaded_Q2}
{{\bm Q}}_{m}
%&={{\bm G}}_2 ~\text{diag} \left({\tilde{\bm d}}_{m}\right)\notag\\
&=\underbrace{{{\bm G}}_2 ~\text{diag} \left({\tilde{\bm u}}\right)}_{{\tilde{\bm R}} } \cdot \underbrace{\text{diag} \left({\tilde{\bm u}}\right)^{-1}\text{diag} \left({\tilde{\bm d}}_{m}\right)}_{\text{diag} \left({{\bm a}}_{m}\right)}
\end{align}
%with $m=1,\ldots, M_1$,
where $\text{diag} \left({{\bm a}}_{m}\right)$ is the diagonal matrix normalized by ${\tilde{\bm u}}$, with ${{\bm a}}_{m}=\text{diag}\left({\tilde{\bm u}}\right)^{-1}{\tilde{\bm d}}_{m} \in {\mathbb{C}^{M_2\times 1}}, \forall m=1,\ldots, M_1$ being the scaling vector.
By 
substituting ${{\bm Q}}_{m}$ of \eqref{cascaded_Q2} into \eqref{superposed3}, the channel model in \eqref{superposed3} can be rewritten as
\vspace{-0.2cm}
\begin{align}
{\bm h}
%&=\sum_{m=1}^{M_1} {\tilde{\bm R}}~\text{diag} \left({{\bm a}}_{m}\right) {\bm \theta}_2 {\theta}_{1,m} +{\tilde{\bm R}}{\bm \theta}_2+{{\bm R}}{\bm \theta}_1\notag\\
&=\sum_{m=1}^{M_1} {\tilde{\bm R}}~\text{diag} \left({{\bm a}}_{m}\right) {\bm \theta}_2 {\theta}_{1,m} 
+{\tilde{\bm R}}{\bm \theta}_2
+{{\bm R}}{\bm \theta}_1
\label{superposed4}.
\end{align}
%According to \eqref{superposed4}, it is sufficient to acquire the CSI of ${{\bm R}}$, ${\tilde{\bm R}}$, and the normalized vectors $\left\{{{\bm a}}_{m}\right\}_{m=1}^{M_1}$, for which the number of channel coefficients to be estimated is substantially reduced as compared to that in \eqref{superposed3}.
%Based on the relationship disclosed in \eqref{cascaded_Q2}, we first propose a new channel estimation scheme with the simple ON/OFF IRS reflection (for decoupling reflection links in different phases during the channel estimation stage), whose main procedures are described as follows and will be further elaborated in the rest of this subsection. 
%\begin{itemize}
%	\item With all the subsurfaces at IRS~2 turned OFF (i.e., ${\bm \theta}_2={\bf 0}_{M_2 \times 1}$), we estimate 
%	the CSI of ${{\bm R}}$ by varying the reflection of IRS~1 over different time slots;
%	\item With all the subsurfaces at IRS~1 turned OFF (i.e., ${\bm \theta}_1={\bf 0}_{M_1 \times 1}$), 
%	we estimate the CSI of $ {\tilde{\bm R}}$  by varying the reflection of IRS~2 over different time slots; 
%	\item  With a portion of $M_1$ subsurfaces at IRS~1 turned ON sequentially, we estimate the CSI of $\left\{{{\bm a}}_{m}\right\}_{m=1}^{M_1}$ in the remaining time slots, and recover each ${{\bm Q}}_{m}$ from the estimated ${\tilde{\bm R}}$ and ${{\bm a}}_{m}$ according to \eqref{cascaded_Q2}.
%\end{itemize}
According to \eqref{superposed4}, it is sufficient to acquire the CSI of ${{\bm R}}$, ${\tilde{\bm R}}$, and the scaling vectors $\left\{{{\bm a}}_{m}\right\}_{m=1}^{M_1}$ for designing the passive beamforming for data transmission in the double-IRS assisted single-user system.
Based on the channel relationship disclosed in \eqref{cascaded_Q2}, 
we propose to decouple the channel estimation for the single- and double-reflection links into three phases, for which the  main procedures are described as follows and will be elaborated in the subsequent subsections.
\begin{itemize}
	\item \emph{Estimation of the two single-reflection channels $\left\{{{\bm R}},{\tilde{\bm R}}\right\}$:} With all the subsurfaces at IRS~2 (IRS~1) turned OFF, the BS estimates
	${{\bm R}}$ (${\tilde{\bm R}}$) based on the time-varying training reflection of IRS~1 (IRS~2) and the pilot symbols sent by the user;
	%	\item With all the subsurfaces at IRS~1 turned OFF (i.e., ${\bm \theta}_1={\bf 0}_{M_1 \times 1}$), 
	%	we estimate the CSI of ${\tilde{\bm R}}$  by varying the reflection of IRS~2 over different time slots; 
	\item  \emph{Estimation of the double-reflection channel $\left\{ {{\bm Q}}_{m} \right\}_{m=1}^{M_1}$:} After canceling the signals over the two single-reflection links and taking the estimated ${\tilde{\bm R}}$ as the reference CSI,
	the BS estimates $\left\{{{\bm a}}_{m}\right\}_{m=1}^{M_1}$ for the double-reflection link.
\end{itemize}
%	the CSI of $\left\{{{\bm a}}_{m}\right\}_{m=1}^{M_1}$ is estimated either by turning ON one out of $M_1$ subsurfaces at IRS~1 sequentially or varying the full-ON reflections over different time slots. The CSI of each ${{\bm Q}}_{m}$ is then obtained from the estimated ${\tilde{\bm R}}$ and ${{\bm a}}_{m}$ according to \eqref{cascaded_Q2}.
\vspace{-0.5cm}
\subsection{Phase I: Estimation of ${{\bm R}}$}\label{Pro_R1}
%{\bf Phase I (Estimation of ${{\bm R}}$):}
With all the subsurfaces at IRS~2 turned OFF (i.e., ${\bm \theta}_{2,\rm I}^{(i)}={\bm 0}_{M_2 \times 1},\forall i$) in Phase~I, the channel model in \eqref{superposed4} reduces to the single-reflection channel related to IRS 1 only. In this case, the received
signal of the BS at time slot $i$ of Phase~I can be expressed as
\vspace{-0.2cm}
\begin{align}
{\bm y}_{\rm I}^{(i)}&= {{\bm R}}~ {\bm \theta}_{1,\rm I}^{(i)} x_{\rm I}^{(i)} +{\bm v}_{\rm I}^{(i)}, \quad i=1,\ldots,I_1\label{rec_IRS1}
\end{align}
where $I_1$ denotes the number of pilot symbols in Phase I,
$x_{\rm I}^{(i)}$ represents the pilot symbol transmitted by the user which is simply set as $x_{\rm I}^{(i)}=1$ for ease of exposition, and ${\bm v}_{\rm I}^{(i)}\sim {\mathcal N_c }({\bm 0}, \sigma^2{\bm I}_N)$ is the additive white Gaussian noise (AWGN) vector at the BS with $\sigma^2$ being the normalized noise power.
% with normalized noise power of $\sigma^2$ at the BS.  
By stacking the received signal vectors $\{{\bm y}_{\rm I}^{(i)}\}_{i=1}^{I_1}$ into ${\bm Y}_{\rm I}=\left[{\bm y}_{\rm I}^{(1)},\ldots,{\bm y}_{\rm I}^{(I_1)}\right]$, we obtain
\vspace{-0.2cm}
\begin{align}\label{Rec_P1}
{\bm Y}_{\rm I}={{\bm R}} {\bm \Theta}_{1,\rm I} + {\bm V}_{\rm I}
\end{align}
where ${\bm \Theta}_{1,\rm I}=\left[{\bm \theta}_{1,\rm I}^{(1)},\ldots,{\bm \theta}_{1,\rm I}^{(I_1)}\right]$ denotes the training reflection matrix at IRS~1 in Phase I
% that collects all training reflection coefficients $\{{\bm \theta}_{1,\rm I}^{(i)}\}_{i=1}^{I_1}$ 
and ${\bm V}_{\rm I}=\left[{\bm v}_{\rm I}^{(1)},\ldots,{\bm v}_{\rm I}^{(I_1)}\right]\in {\mathbb{C}^{N\times I_1}}$ denotes the corresponding AWGN matrix. By properly constructing the training
reflection matrix of IRS~1 such that ${\rm rank}\left( {\bm \Theta}_{1,\rm I} \right)=M_1$, the least-square (LS) estimate of ${{\bm R}}$ based on \eqref{Rec_P1} is given by
\vspace{-0.15cm}
\begin{align}\label{Est_P1}
{\hat{\bm R}}={\bm Y}_{\rm I}  {\bm \Theta}_{1,\rm I}^H\left({\bm \Theta}_{1,\rm I}{\bm \Theta}_{1,\rm I}^H\right)^{-1}.
\end{align}
Moreover, $I_1\ge M_1$ is required to satisfy ${\rm rank}\left( {\bm \Theta}_{1,\rm I} \right)=M_1$, which can be achieved via
different IRS training reflection designs (e.g., the ON/OFF based design \cite{yang2019intelligent} or the more efficient  orthogonal matrix-based designs \cite{zheng2019intelligent,zheng2020intelligent,zheng2020fast,you2019progressive}).
%with ${\bm \Theta}_{1,\rm I}^\dag={\bm \Theta}_{1,\rm I}^H\left({\bm \Theta}_{1,\rm I}{\bm \Theta}_{1,\rm I}^H\right)^{-1}$.
\vspace{-0.4cm}
\subsection{Phase II: Estimation of ${\tilde{\bm R}}$}\label{Pro_R2}
%{\bf Phase II (Estimation of ${\tilde{\bm R}}$):}
With all the subsurfaces at IRS~1 turned OFF (i.e., ${\bm \theta}_{1,\rm II}^{(i)}={\bm 0}_{M_1 \times 1},\forall i$) in Phase II, the channel model in \eqref{superposed4} reduces to the single-reflection channel related to IRS 2 only.
In this case, with $x_{\rm II}^{(i)}=1,\forall i$ being the pilot symbol transmitted by the user, the received
signal matrix of the BS over $I_2$ pilot symbols of Phase~II is similarly obtained as
\vspace{-0.2cm}
\begin{align}\label{Rec_P2}
{\bm Y}_{\rm II}={\tilde{\bm R}} {\bm \Theta}_{2,\rm II} + {\bm V}_{\rm II}
\end{align}
where ${\bm \Theta}_{2,\rm II}=\left[{\bm \theta}_{2,\rm II}^{(1)},\ldots,{\bm \theta}_{2,\rm II}^{(I_2)}\right]$ is the training reflection matrix at IRS~2 in Phase II and ${\bm V}_{\rm II}\in {\mathbb{C}^{N\times I_2}}$
is the corresponding AWGN matrix.
%Based on the channel model in \eqref{superposed4} with all the subsurfaces at IRS~1 turned OFF (i.e., ${\bm \theta}_1^{(i)}={\bm 0}_{M_1 \times 1},\forall i$) and following the similar procedures in Section~\ref{Pro_R1},
Accordingly, the LS estimate of ${\tilde{\bm R}}$ based on \eqref{Rec_P2} is given by
\vspace{-0.2cm}
\begin{align}\label{Est_P2}
{\hat{\tilde{\bm R}}}={\bm Y}_{\rm II}  {\bm \Theta}_{2,\rm II}^H\left({\bm \Theta}_{2,\rm II}{\bm \Theta}_{2,\rm II}^H\right)^{-1}.
%={\tilde{\bm R}}+{\bm V}_{\rm II}{\bm \Theta}_{2,\rm II}^\dag
\end{align}
%with ${\bm Y}_{\rm II}={\tilde{\bm R}} {\bm \Theta}_{2,\rm II} + {\bm V}_{\rm II}$ denoting the received signal matrix at the BS and $x_{\rm II}^{(i)}=1$ being the pilot symbol transmitted by the user,
 Similarly, $I_2\ge M_2$ is required to achieve ${\rm rank}\left( {\bm \Theta}_{2,\rm II} \right)=M_2$ via different IRS training reflection designs \cite{yang2019intelligent,zheng2019intelligent,zheng2020intelligent,zheng2020fast,you2019progressive}.

\vspace{-0.4cm}
\subsection{Phase III: Estimation of $\{{{\bm a}}_{m}\}_{m=1}^{M_1}$}
%{\bf Phase III (Estimation of $\left\{{{\bm a}}_{m}\right\}_{m=1}^{M_1}$):} 
With the estimated $\left\{{{\bm R}},{\tilde{\bm R}}\right\}$ in Phases I and II, we further estimate each scaling vector ${{\bm a}}_{m}$ for the double-reflection link.
%to obtain the CSI of each ${{\bm Q}}_{m}$ according to \eqref{cascaded_Q2}.
At time slot $i$ of Phase III with $x_{\rm III}^{(i)}=1,\forall i$ being the pilot symbol transmitted by the user, the received signal at the BS based on the channel model in \eqref{superposed4} can be expressed as
\vspace{-0.2cm}
\begin{align}\label{Rec_P3_1}
&{\bm y}_{\rm III}^{(i)}\hspace{-0.1cm}=\hspace{-0.2cm} \sum_{m=1}^{M_1} \hspace{-0.1cm}{\tilde{\bm R}} \text{diag} \hspace{-0.1cm}\left({{\bm a}}_{m}\right) {\bm \theta}_{2,\rm III}^{(i)} {{\theta}}_{1,m,{\rm III}}^{(i)}\hspace{-0.1cm}+\hspace{-0.1cm}{\tilde{\bm R}} {\bm \theta}_{2,\rm III}^{(i)}
\hspace{-0.1cm}+\hspace{-0.1cm} {{\bm R}}{\bm \theta}_{1,\rm III}^{(i)} \hspace{-0.1cm}+\hspace{-0.1cm} {\bm v}_{\rm III}^{(i)}\notag\\
&=\hspace{-0.15cm}\sum_{m=1}^{M_1}\hspace{-0.1cm} {{\theta}}_{1,m,{\rm III}}^{(i)}{\tilde{\bm R}}\text{diag} \hspace{-0.1cm}\left({\bm \theta}_{2,\rm III}^{(i)}\right)\hspace{-0.1cm}{{\bm a}}_{m}  \hspace{-0.1cm}+\hspace{-0.1cm}{\tilde{\bm R}} {\bm \theta}_{2,\rm III}^{(i)}
\hspace{-0.1cm}+\hspace{-0.1cm} {{\bm R}}{\bm \theta}_{1,\rm III}^{(i)} \hspace{-0.1cm}+\hspace{-0.1cm} {\bm v}_{\rm III}^{(i)}\hspace{-0.1cm}
\end{align}
where ${\bm v}_{\rm III}^{(i)}\sim {\mathcal N_c }({\bm 0}, \sigma^2{\bm I}_N)$ is the AWGN vector. Given the estimated $\left\{{{\bm R}},{\tilde{\bm R}}\right\}$, the pilot signals over the two single-reflection links can be removed from \eqref{Rec_P3_1}, and thus the effective received signal over the double-reflection link at the BS is given by\footnote{For ease of exposition, we assume perfect cancellation of the pilot signals over the two single-reflection links; while the residual interference due to imperfect cancellation with estimated $\left\{{{\bm R}},{\tilde{\bm R}}\right\}$ will be taken into account for evaluating the channel estimation performance via simulations in Section~\ref{Sim}.}
\vspace{-0.2cm}
\begin{align}\label{Rec_P3_2}
{\bar {\bm y}}_{\rm III}^{(i)}&={\bm y}_{\rm III}^{(i)}-{\tilde{\bm R}} {\bm \theta}_{2,\rm III}^{(i)}
- {{\bm R}}{\bm \theta}_{1,\rm III}^{(i)} \notag\\
&= \sum_{m=1}^{M_1} {{\theta}}_{1,m,{\rm III}}^{(i)} {\tilde{\bm R}}~\text{diag} \left({\bm \theta}_{2,\rm III}^{(i)}\right){{\bm a}}_{m}  + {\bm v}_{\rm III}^{(i)}.
\end{align}
For the estimation of $\left\{ {{\bm a}}_{m}\right\}_{m=1}^{M_1}$, we consider the following two cases.

{\bf Case 1:} $N\ge M_2$. In this case, we consider the fixed full-ON reflection of IRS~2 (say, ${\bm \theta}_{2,\rm III}^{(i)}={\bm 1}_{M_2 \times 1}, \forall i$) and turn ON
one out of the $M_1$ subsurfaces at IRS~1 (say, ${\theta}_{1,i,{\rm III}}^{(i)}=1$ and ${\theta}_{1,m,{\rm III}}^{(i)}=0, \forall m\neq i$) sequentially to estimate each ${{\bm a}}_{i}$, for which the received signal of the BS at time slot $i$ of Phase III is given by
\vspace{-0.2cm}
\begin{align}\label{Rec_P3c1}
{\bar {\bm y}}_{\rm III}^{(i)}=  {\tilde{\bm R}} {{\bm a}}_{i} 
+ {\bm v}_{\rm III}^{(i)},
\quad i=1,\ldots,M_1.
\end{align}
%where ${\bm \theta}_{2,\rm III}^{(i)}={\bm \theta}_{2,\rm III},\forall i$ is assumed without loss generality.
%Specifically, based on channel model in \eqref{superposed4} with one out of $M_1$ subsurfaces (say, subsurface $i$) at IRS~1 turned ON (i.e., ${\theta}_{1,i}=1$ and ${\theta}_{1,m,{\rm III}}=0, \forall m\neq i$) sequentially, the received signal at the BS can be expressed as 
%\vspace{-0.2cm}\begin{align}
%{\bar {\bm y}}_{\rm III}^{(i)}=  {\theta}_{1,i} {\tilde{\bm R}}~\text{diag} \left({\bm \theta}_{2,\rm III}\right) {{\bm a}}_{i} 
%+{\tilde{\bm R}} {\bm \theta}_{2,\rm III} +{{\bm r}}_{i}{\theta}_{1,i}+{\bar{\bm v}}_k^{(i)},
%\quad i=1,\ldots,M_1
%\end{align}
%where ${{\bm r}}_{i}$ is the $i$-th column of ${{\bm R}}$ and
%${\bar{\bm v}}_k^{(i)}$ denotes the additive
%white Gaussian noise (AWGN) vector at the BS with variance $\sigma^2$.
%As the CSI of ${{\bm R}}$ and $ {\tilde{\bm R}}$ has been acquired, by setting ${\bm \theta}_{2,\rm III}={\bm 1}_{M_2\times 1}$ and ${\theta}_{1,i}=1$ for simplicity, 
Based on \eqref{Rec_P3c1}, the LS estimate of each ${{\bm a}}_{i}$ is given by
\vspace{-0.2cm}
\begin{align}\label{est_a}
{\hat{\bm a}}_{i}&=\left( {\tilde{\bm R}}^H{\tilde{\bm R}}\right)^{-1}{\tilde{\bm R}}^H{\bar {\bm y}}_{\rm III}^{(i)}, \quad i=1,\ldots,M_1.
%={{\bm a}}_{i}+{\theta}_{1,i}^H\left({\tilde{\bm R}}~\text{diag} \left({\bm \theta}_{2,\rm III}\right)\right)^{\dagger}{\bm v}_{\rm III}^{(i)}\quad i=1,\ldots,M_1.
\end{align} 
%where $\left({\tilde{\bm R}}~\text{diag} \left({\bm \theta}_{2,\rm III}\right)\right)^{\dagger}=\left(\text{diag} \left({\bm \theta}_{2,\rm III}\right)^H{\tilde{\bm R}}^H{\tilde{\bm R}}~\text{diag} \left({\bm \theta}_{2,\rm III}\right)\right)^{-1}\text{diag} \left({\bm \theta}_{2,\rm III}\right)^H{\tilde{\bm R}}^H$.
%Therefore, the minimum training overhead in terms of time slots is \rev{$2M_1+M_2$} in total for the channel estimation with ON/OFF IRS reflection.

{\bf Case 2:} $N< M_2$. 
%Following the same procedures of Case 1, the minimum number of time slots required to estimate each ${{\bm a}}_{m}$ is $\left\lceil\frac{M_2}{N}\right\rceil$ and thus the total training overhead for Phase~II is $M_1\left\lceil\frac{M_2}{N}\right\rceil$. However, the training overhead can be reduced to $\left\lceil\frac{M_1 M_2}{N}\right\rceil$ if we jointly estimate $\left\{{{\bm a}}_{m}\right\}$.
%by first dividing $\left\{{{\bm a}}_{m}\right\}_{m=1}^{M_1}$ into $L=\left\lceil\frac{M_1}{N}\right\rceil$ groups and then jointly estimating ${{\bm a}}_{m}$'s in each group, which can follow the similar procedures to \cite{wang2019channel} and thus is omitted for brevity.
In this case, we cannot estimate each ${{\bm a}}_{m}$ separately according to \eqref{est_a}, since ${\tilde{\bm R}}$ is (column) rank-deficient, i.e., ${\rm rank}\left( {\tilde{\bm R}} \right)=N < M_2$.
Alternatively, we consider the joint estimation of $\{{{\bm a}}_{m}\}_{m=1}^{M_1}$ by stacking the received signal vectors $\{ {\bar {\bm y}}_{\rm III}^{(i)}\}$ in \eqref{Rec_P3_2} over $I_3$ time slots of Phase III, which is given by
% as ${\bar {\bm y}}_{\rm III}=\left[({\bar {\bm y}}_{\rm III}^{(1)})^T,\ldots,({\bar {\bm y}}_{\rm III}^{(I_3)})^T\right]^T$, we obtain
%\vspace{-0.2cm}\begin{align}\label{Rec_P3_3}
%{\bar {\bm y}}_{\rm III}=
%\underbrace{\begin{bmatrix}
%	{{\theta}}_{1,1}^{(1)}{\tilde{\bm R}} ~\text{diag} \left({\bm \theta}_{2,\rm III}^{(1)}\right) &\cdots &{{\theta}}_{1,M_1}^{(1)}{\tilde{\bm R}} ~\text{diag} \left({\bm \theta}_{2,\rm III}^{(1)}\right)\\ 
%	\vdots &&\vdots\\ 
%	{{\theta}}_{1,1}^{(I_3)}{\tilde{\bm R}} ~\text{diag} \left({\bm \theta}_{2,\rm III}^{(I_3)}\right) &\cdots &{{\theta}}_{1,M_1}^{(I_3)}{\tilde{\bm R}} ~\text{diag} \left({\bm \theta}_{2,\rm III}^{(I_3)}\right)
%	\end{bmatrix}}_{ {\bm C}\in {\mathbb{C}^{ I_3N  \times M_1M_2 }}}
%\underbrace{\begin{bmatrix}
%	{{\bm a}}_{1}\\\vdots\\{{\bm a}}_{M_1}
%	\end{bmatrix}}_{ {\bar{\bm a}} }
%+ \begin{bmatrix}
%{\bm v}_{\rm III}^{(1)}\\\vdots\\{\bm v}_{\rm III}^{(I_3)}
%\end{bmatrix}
%\end{align}
\vspace{-0.1cm}
\begin{align}\label{Rec_P3_3}
\hspace{-0.25cm}
\underbrace{\begin{bmatrix}
{\bar {\bm y}}_{\rm III}^{(1)}\\\vdots\\{\bar {\bm y}}_{\rm III}^{(I_3)}
\end{bmatrix}}_{{\bar {\bm y}}_{\rm III}}
\hspace{-0.15cm}=\hspace{-0.15cm}
\underbrace{\begin{bmatrix}
	\hspace{-0.1cm}\left({\bm \theta}_{1,\rm III}^{(1)}\right)^T \hspace{-0.1cm}\otimes\hspace{-0.1cm} {\tilde{\bm R}} ~\text{diag} \left({\bm \theta}_{2,\rm III}^{(1)}\right) \\ 
	\vdots \\ 
	\hspace{-0.1cm}\left({\bm \theta}_{1,\rm III}^{(I_3)}\right)^T \hspace{-0.1cm}\otimes\hspace{-0.1cm} {\tilde{\bm R}} ~\text{diag} \left({\bm \theta}_{2,\rm III}^{(I_3)}\right)
	\end{bmatrix}}_{ {\bm C}\in {\mathbb{C}^{ I_3N  \times M_1M_2 }}}
\underbrace{\begin{bmatrix}
	{{\bm a}}_{1}\\\vdots\\{{\bm a}}_{M_1}
	\end{bmatrix}}_{ {{\bm \eta}} }
\hspace{-0.1cm}+ \hspace{-0.1cm}\begin{bmatrix}
{\bm v}_{\rm III}^{(1)}\\\vdots\\{\bm v}_{\rm III}^{(I_3)}
\end{bmatrix}\hspace{-0.1cm}.\hspace{-0.1cm}
\end{align}
%for which the reflections of IRSs~1 and 2 varies over $I_3$ time slots in Phase III.
As such, by properly designing the training
reflection coefficients $\left\{{\bm \theta}_{1,\rm III}^{(i)}\right\}_{i=1}^{I_3}$ and $\left\{{\bm \theta}_{2,\rm III}^{(i)}\right\}_{i=1}^{I_3}$ of IRSs~1 and 2
such that ${\rm rank}\left( {\bm C} \right)=M_1M_2$,
the LS estimate of ${{\bm \eta}}$ can be obtained as
\vspace{-0.1cm}
\begin{align}\label{est_a2}
{\hat{{\bm \eta}}}=\left({\bm C}^H{\bm C}\right)^{-1}{\bm C}^H {\bar {\bm y}}_{\rm III}.
\end{align}
Note that $I_3N  \ge M_1M_2$ is the necessary condition to achieve ${\rm rank}\left( {\bm C} \right)=M_1M_2$ with $I_3$ being an integer, and thus we have $I_3 \ge \left\lceil\frac{M_1 M_2}{N}\right\rceil$. Moreover, we can design the training
reflections of IRSs~1 and 2 based on the orthogonal matrices as in \cite{zheng2019intelligent,zheng2020intelligent,zheng2020fast,you2019progressive}.
% to exploit the large aperture gain of IRSs so as to improve the channel estimation accuracy.
It is worth pointing out that the channel estimation based on \eqref{Rec_P3_3} and \eqref{est_a2} 
with the orthogonal matrix-based training design
can also be applied to the case of $N< M_2$ to achieve better channel estimation performance at the expense of higher complexity due to the higher-dimensional matrix inversion operation and the joint estimation of $\{{{\bm a}}_{m}\}_{m=1}^{M_1}$.

Finally, with the estimated CSI of ${\tilde{\bm R}}$ and $\left\{{{\bm a}}_{m}\right\}_{m=1}^{M_1}$, we can obtain the estimated CSI of $\left\{ {{\bm Q}}_{m} \right\}_{m=1}^{M_1}$ according to \eqref{cascaded_Q2}.
\vspace{-0.25cm}
\section{Channel Estimation for Multi-User Case}\label{MU_ON/OFF}
For the multi-user channel estimation, a straightforward method is by adopting the single-user channel estimation design in 
Section~\ref{ON/OFF} to estimate the channels of $K$ users separately over consecutive time, which, however, increases the total training overhead by $K$ times as compared to the single-user case and thus is practically prohibitive if $K$ is large. To reduce the overall training overhead, we extend the channel estimation scheme for the single-user case in Section~\ref{ON/OFF}
to the general multi-user case in this section. 
%Similar to \cite{wang2019channel,zheng2020intelligent}, the estimated CSI of an arbitrary user (referred to as the reference user) is exploited to substantially reducing the channel training overhead of the other users.
By exploiting the fact that the other users' cascaded channels are scaled versions of the cascaded channel of an arbitrary user (referred to as the reference user) \cite{wang2019channel,zheng2020intelligent}, the channel training overhead can be substantially reduced, which is elaborated in the following.

After estimating an arbitrary user's cascaded channel as in the single-user case of Section~\ref{ON/OFF}, the  cascaded channels of the other users can be efficiently obtained by exploiting the fact that all the users
share the common IRS~2$\rightarrow$BS (i.e, ${{\bm G}}_2$), IRS~1$\rightarrow$BS (i.e, ${{\bm G}}_1$), and IRS~1$\rightarrow$IRS~2 (i.e, ${\bm D}$) links in \eqref{superposed0} in their respective single- and double-reflection channels.
In particular, if given the cascaded CSI of any user (say, ${{\bm R}}_{1}$, ${\tilde{\bm R}}_{1}$, and $\left\{{{\bm Q}}_{1,m}\right\}_{m=1}^{M_1}$ of user 1) as the reference CSI,
we can rewrite 
the two single-reflection channels $\{{{\bm R}}_{k},{\tilde{\bm R}}_{k}\}$ in \eqref{superposed3} as
\begin{align}
{{\bm R}}_{k}&
%={{\bm G}}_1   \text{diag} \left( {{\bm u}}_{k} \right)
={{\bm G}}_1\text{diag} \left({{\bm u}}_{1}\right) \cdot \text{diag}\left({{\bm u}}_{1}\right)^{-1} {{\bm u}}_{k}=
{{\bm R}}_{1}\text{diag} \left({{\bm b}}_{k} \right)\label{R_k1}\\
{\tilde{\bm R}}_{k}&
%={{\bm G}}_2   \text{diag} \left( {\tilde{\bm u}}_{k} \right)
={{\bm G}}_2\text{diag} \left({\tilde{\bm u}}_{1}\right) \cdot \text{diag}\left({\tilde{\bm u}}_{1}\right)^{-1} {\tilde{\bm u}}_{k}=
{\tilde{\bm R}}_{1} \text{diag} \left( {\tilde {\bm b}}_{k} \right)\label{R_k2}
\end{align}
and the double-reflection channel $\left\{{{\bm Q}}_{k,m}\right\}_{m=1}^{M_1}$ in \eqref{cascaded_Q} as
%\vspace{-0.2cm}
\begin{align}\label{cascaded_Q4}
{{\bm Q}}_{k,m}=&{{\bm G}}_2 \text{diag} \left({\tilde{\bm d}}_{k,m}\right)=
{{\bm G}}_2 \text{diag} \left({\bm d}_{m} {{u}}_{k,m} \right)\notag\\
=&{{\bm G}}_2 \text{diag} \left({\bm d}_{m} {{u}}_{1,m}\right) \cdot{{u}}_{1,m}^{-1} {{u}}_{k,m}= {{\bm Q}}_{1,m}{{b}}_{k,m}
\end{align}
%${\bm d}_{m}$ is the $m$-th column of ${\bm D}$, and 
where ${{\bm b}}_{k}\triangleq\left[{{b}}_{k,1},\ldots,{{b}}_{k,M_1}\right]^T=\text{diag} \left({{\bm u}}_{1}\right)^{-1} {{\bm u}}_{k}$ and
${\tilde {\bm b}}_{k}=\text{diag} \left({\tilde{\bm u}}_{1}\right)^{-1} {\tilde{\bm u}}_{k}$
are the user~$k$$\rightarrow$IRS~1 and user~$k$$\rightarrow$IRS~2 channel vectors normalized by ${{\bm u}}_{1}$ and ${\tilde{\bm u}}_{1}$, respectively.
By substituting \eqref{R_k1}-\eqref{cascaded_Q4} into \eqref{superposed3}, we can re-express the channel model in \eqref{superposed3} as 
\vspace{-0.2cm}
\begin{align}
&\hspace{-0.2cm}{\bm h}_k
=\hspace{-0.1cm}\sum_{m=1}^{M_1}\hspace{-0.1cm} {{\bm Q}}_{1,m} {\bm \theta}_2 {\theta}_{1,m} {{b}}_{k,m}
\hspace{-0.1cm}+\hspace{-0.1cm}{\tilde{\bm R}}_{1} \text{diag} \left( {\tilde {\bm b}}_{k} \right){\bm \theta}_2
\hspace{-0.1cm}+\hspace{-0.1cm}{{\bm R}}_{1}\text{diag} \left({{\bm b}}_{k} \right){\bm \theta}_1\notag\\
&\hspace{-0.3cm}=\hspace{-0.1cm}\sum_{m=1}^{M_1} \hspace{-0.1cm}{{\bm Q}}_{1,m} {\bm \theta}_2 {\theta}_{1,m} {{b}}_{k,m}
\hspace{-0.1cm}+\hspace{-0.1cm}{\tilde{\bm R}}_{1} \text{diag}\hspace{-0.1cm} \left({\bm \theta}_2 \right)  {\tilde {\bm b}}_{k}
\hspace{-0.1cm}+\hspace{-0.1cm}{{\bm R}}_{1}\text{diag}\hspace{-0.1cm} \left({\bm \theta}_1\right){{\bm b}}_{k}
\label{superposed7}.\hspace{-0.2cm}
\end{align}
As such, after acquiring the cascaded CSI of user~1 as in Section~\ref{ON/OFF}, we only need to further estimate $\{{{\bm b}}_{k}\}_{k=2}^K$ and $\{{\tilde {\bm b}}_{k}\}_{k=2}^K$
for the remaining $K-1$ users according to \eqref{R_k1}-\eqref{superposed7}.
In the following, we propose the decoupled channel estimation for $\{{{\bm b}}_{k}\}_{k=2}^K$ and $\{{\tilde {\bm b}}_{k}\}_{k=2}^K$ in Phases IV and V following Phases I-III for estimating the cascaded CSI of the single user (i.e., user 1) in Section \ref{ON/OFF}.
\vspace{-0.3cm}
\subsection{Phase IV: Estimation of $\{{{\bm b}}_{k}\}_{k=2}^K$}\label{pro_b1}
 %with all the subsurfaces at IRS~2 turned OFF and 
 In this phase, we turn OFF all the subsurfaces at IRS~2.
 Based on the channel model in \eqref{superposed7} and denoting $\left\{x_k^{(i)}\right\}_{k=2}^K$ as pilot symbols transmitted by the remaining $K-1$ users, the received signal of the BS at time slot $i$ of Phase IV can be expressed as
\vspace{-0.2cm}
\begin{align}\label{rec_IRS1_MU}
{\bm y}_{\rm IV}^{(i)}&
 %=\sum_{k=2}^{K}  {{\bm R}}_{1}\text{diag} \left({{\bm b}}_{k} \right)~ {\bm \theta}_{1,\rm IV}^{(i)} x_k^{(i)} +{\bm v}^{(i)}
 = \sum_{k=2}^{K} x_k^{(i)} {{\bm R}}_{1} \text{diag} \left( {\bm \theta}_{1,\rm IV}^{(i)}  \right) {{\bm b}}_{k} +{\bm v}_{\rm IV}^{(i)} 
%{\tilde {\bm y}}_k^{(i)}&= {\tilde{\bm R}}_{k}~ {\bm \theta}_2^{(i)} {\tilde x}_k^{(i)} +{\tilde {\bm v}}_k^{(i)}, \quad i=1,\ldots,I_2\label{rec_IRS2}
\end{align}
with ${\bm v}_{\rm IV}^{(i)}\sim {\mathcal N_c }({\bm 0}, \sigma^2{\bm I}_N)$ being the AWGN vector.
%which reduces to the single IRS channel estimation problem for the multi-user case. 
%By exploiting the fact that other users' cascaded channels are scaled versions of use 1's cascaded CSI \cite{wang2019channel,zheng2020intelligent}, two cases are considered.
For the estimation of $\{{{\bm b}}_{k}\}_{k=2}^K$, we consider the following two cases.

{\bf Case 1:} $N\ge M_1$. In this case, the remaining $K-1$ users send pilot symbols sequentially for the BS to estimate each 
${{\bm b}}_{k}$ with $k=2,\ldots,K$. Specifically, with $x_{i+1}^{(i)}=1$ and $x_k^{(i)}=0, \forall k\neq i+1$ 
and the fixed full-ON reflection of IRS~1 (say, ${\bm \theta}_{1,\rm IV}^{(i)}={\bm 1}_{M_1 \times 1}, \forall i$) in Phase IV, the received signal in \eqref{rec_IRS1_MU} 
can be rewritten as
 \vspace{-0.2cm}
 \begin{align}\label{rec_y}
 {\bm y}_{\rm IV}^{(i)}={{\bm R}}_{1} {{\bm b}}_{i+1} +{\bm v}_{\rm IV}^{(i)}, \quad i=1,\ldots,K-1
  \end{align}
and the LS estimate of ${{\bm b}}_{i+1}$ is thus given by
 \vspace{-0.2cm}
 \begin{align}\label{LS_b}
 {\hat{\bm b}}_{i+1}=\left({{\bm R}}_{1}^H{{\bm R}}_{1}\right)^{-1}{{\bm R}}_{1}^H{\bm y}_{\rm IV}^{(i)}, \quad i=1,\ldots,K-1.
 \end{align}
 
 {\bf Case 2:} $N< M_1$. In this case, since ${{\bm R}}_{1}$ is not of full-column rank, i.e., ${\rm rank}\left( {{\bm R}}_{1} \right)=N < M_1$, we cannot estimate each ${{\bm b}}_{k}$ separately according to \eqref{rec_y}.
 As such, we consider the joint estimation of $\{{{\bm b}}_{k}\}_{k=2}^K$ with concurrent pilot symbols sent by
 the remaining $K-1$ users, for which the received signal vector at the BS over $I_4$ pilot symbols is given by
% the remaining $K-1$ users simultaneously send pilot symbols for the joint estimation of $\{{{\bm b}}_{k}\}_{k=2}^K$ over $I_4$ time slots, for which the receive signal at the BS is given by
 \vspace{-0.2cm}
 \begin{align}\label{Rec_P4}
\hspace{-0.2cm} \underbrace{\begin{bmatrix}
 {\bm y}_{\rm IV}^{(1)}\\\vdots\\{\bm y}_{\rm IV}^{(I_4)}
 \end{bmatrix}}_{{\bm y}_{\rm IV}}\hspace{-0.15cm}=\hspace{-0.15cm}
 \underbrace{\begin{bmatrix}
 	\hspace{-0.2cm}\left({\bm x}_{\rm IV}^{(1)}\right)^T \hspace{-0.15cm}\otimes {\bm R} ~\text{diag} \left({\bm \theta}_{1,\rm IV}^{(1)}\right) \\ 
 	\vdots \\ 
 	\hspace{-0.1cm}\left({\bm x}_{\rm IV}^{(I_4)}\right)^T \hspace{-0.15cm}\otimes {\bm R} ~\text{diag} \left({\bm \theta}_{1,\rm IV}^{(I_4)}\right)
 	\end{bmatrix}}_{ {\bm F}\in {\mathbb{C}^{ I_4 N  \times (K-2)M_1 }} }\hspace{-0.1cm}
 \underbrace{\begin{bmatrix}
 	{{\bm b}}_{2}\\\vdots\\{{\bm b}}_{K}
 	\end{bmatrix}}_{ {\bm \lambda} }
 + \begin{bmatrix}
 {\bm v}_{\rm IV}^{(1)}\\\vdots\\{\bm v}_{\rm IV}^{(I_4)}
 \end{bmatrix}
 \end{align}
 where ${\bm x}_{\rm IV}^{(i)}=\left[x_2^{(i)},\ldots, x_K^{(i)}\right]^T$ denotes the pilot symbol vector. As such, by properly designing the training
 reflection coefficients $\left\{{\bm \theta}_{1,\rm IV}^{(i)}\right\}_{i=1}^{I_4}$ of IRS~1 and the pilot symbol vectors $\left\{{\bm x}_{\rm IV}^{(i)}\right\}_{i=1}^{I_4}$
 such that ${\rm rank}\left( {\bm F} \right)=(K-2)M_1$,
 the LS estimate of ${\bm \lambda}$ is given by
 \vspace{-0.2cm}
 \begin{align}\label{est_lamda}
 {\hat{\bm \lambda}}=\left({\bm F}^H{\bm F}\right)^{-1}{\bm F}^H {\bm y}_{\rm IV}.
 \end{align}
 Since $I_4 N \ge (K-2)M_1$ is required to ensure the condition of ${\rm rank}\left( {\bm F} \right)=(K-2)M_1$
 with $I_4$ being an integer, we have $I_4  \ge \left\lceil\frac{(K-2)M_1}{N}\right\rceil$.
 Furthermore, we can construct the training
 reflection coefficients $\left\{{\bm \theta}_{1,\rm IV}^{(i)}\right\}_{i=1}^{I_4}$ of IRS~1 and the pilot symbol vectors $\left\{{\bm x}_{\rm IV}^{(i)}\right\}_{i=1}^{I_4}$ from some orthogonal matrices as in \cite{zheng2019intelligent,zheng2020intelligent,zheng2020fast,you2019progressive} to achieve ${\rm rank}\left( {\bm F} \right)=(K-2)M_1$.
 \vspace{-0.35cm}
 \subsection{Phase V: Estimation of $\{{\tilde {\bm b}}_{k}\}_{k=2}^K$}
 Similarly, based on the channel model in \eqref{superposed7} with all the subsurfaces at IRS~1 turned OFF, the received signal of the BS at time slot $i$ of Phase V can be expressed as
 \vspace{-0.2cm}
 \begin{align}\label{rec_IRS2_MU}
 {\bm y}_{\rm V}^{(i)}&
 = \sum_{k=2}^{K} x_k^{(i)} {\tilde{\bm R}}_{1} \text{diag} \left( {\bm \theta}_{2,\rm V}^{(i)}  \right) {\tilde {\bm b}}_{k} +{\bm v}_{\rm V}^{(i)}
 \end{align}
 with $\left\{x_k^{(i)}\right\}_{k=2}^K$ being pilot symbols transmitted by the remaining $K-1$ users and
 ${\bm v}_{\rm V}^{(i)}\sim {\mathcal N_c }({\bm 0}, \sigma^2{\bm I}_N)$ being the AWGN vector.
As such, following the similar procedures in Section~\ref{pro_b1}, we can estimate $\{{\tilde {\bm b}}_{k}\}_{k=2}^K$ for the two cases of $N\ge M_2$ and $N< M_2$ with minimum training overhead of $K-1$ and $\left\lceil\frac{(K-1) M_2}{N}\right\rceil$, respectively, 
whose details are thus omitted for brevity.

 \vspace{-0.4cm}
\section{Simulation Results}\label{Sim}
In this section, we present simulation results to numerically validate the effectiveness of the proposed channel estimation scheme for the double-IRS assisted multi-user MIMO system.
Under a three-dimensional (3D) Cartesian coordinate system, we assume that the central (reference) points of the BS, IRS~2, IRS~1, and user cluster are located at $(1,0,2)$, $(0,0.5,1)$, $(0,49.5,1)$, and $(1,50,0)$ in meter (m), respectively. 
Moreover, we assume that
the BS is equipped with a uniform linear array (ULA); while the two distributed IRSs are equipped with uniform planar arrays (UPAs).
%The azimuth angles of IRSs 1 and 2 with respect to the $x$-axis are set as $\pi/4$ and $3\pi/4$, respectively.
%The total number of IRS elements is ${\tilde M}=800$ as our budget for the two distributed IRS.
As the element-grouping strategy in \cite{yang2019intelligent,zheng2019intelligent},
each IRS subsurface is a small-size UPA composed of $5\times 5$ adjacent IRS elements that share a common phase shift for reducing design complexity.
% in the double-IRS assisted multi-user MIMO system
%As such, the total number of IRS subsurfaces is $M=32$ as our budget.
The distance-dependent channel path loss is modeled as $\gamma=\gamma_0/ d^\alpha$, where $\gamma_0$ denotes the path loss at the reference distance of 1~m which is set as $\gamma_0=-30$~dB for all individual links, $d$ denotes the individual link distance, and $\alpha$ denotes the path loss exponent which is set as $2.2$ for the link between the user cluster/BS and its nearby serving IRS (due to the short distance) and set as $3$ for the other links (due to the relatively large distance).

Due to the very limited work on channel estimation for the double-IRS assisted system, we extend the channel estimation method proposed in \cite{you2020wireless} as the benchmark scheme for comparison,
where the double-reflection channel is estimated at each BS antenna in parallel without exploiting the (common) channel relationship with the single-reflection channels, and the cascaded channels of $K$ users are separately estimated over consecutive time.
%the estimation of the single-reflection channels follows the same procedures in Section \ref{ON/OFF}
Moreover, as the single-reflection channels were ignored in \cite{you2020wireless}, the same channel estimation procedures for the single-reflection channels in Sections \ref{Pro_R1} and \ref{Pro_R2} are adopted for each user in the benchmark scheme.
%We consider the benchmark scheme proposed in \cite{you2020wireless} without exploiting the channel correlation among the single- and double-reflection channels as well as multiple users, where the estimation of the single-reflection channels follows the same procedures in Section \ref{ON/OFF}.
The channel training overhead comparison between the proposed and benchmark schemes is shown in Table~\ref{Table of estimation}, where $M_1=M_2=M/2$ is assumed for ease of exposition. As can be seen, by exploiting the peculiar channel relationship over double-reflection channels and multiple users, the proposed channel estimation scheme incurs much lower training overhead than the benchmark scheme.
\begin{table}[!t]
	\begin{center}\caption{Channel training overhead comparison}\label{Table of estimation}
		\resizebox{0.4\textwidth}{!}{
			\begin{tabular}{|c|c|c|}
				\hline
				\multirow{2}{*}{}                  & \multicolumn{2}{c|}{Minimum number of pilot symbols} \\ \cline{2-3} 
				& $N\ge M/2$                    & $N< M/2$                    \\ \hline
				Proposed scheme & $\frac{3}{2}M+2(K-1)$ & $M+\left\lceil\frac{M^2}{4N}\right\rceil+2\left\lceil\frac{(K-1) M}{2N}\right\rceil$ \\ \hline
				Benchmark scheme based on \cite{you2020wireless} & \multicolumn{2}{c|}{$KM+\frac{1}{4}KM^2$}                   \\ \hline
			\end{tabular}
		}
	\end{center}
	\vspace{-0.6cm}
\end{table}

In the following simulations, we calculate the normalized mean squared error (MSE) for the single- and double-reflection channels over $1,000$ independent fading channel realizations. For example, the normalized MSE of the cascaded user~$k$$\rightarrow$IRS~1$\rightarrow$BS channel ${{\bm R}}_{k}$ is given by
\vspace{-0.2cm}
\begin{align}
{\bar\varepsilon}=\frac{1}{KNM_1}  \sum_{k=1}^K    {\mathbb E}\left\{\left\| {\hat{\bm R}}_{k}-{{\bm R}}_{k}\right\|^{2}_F \Big/{ \left\|{{\bm R}}_{k}\right\|^{2}_F} \right\}.
\end{align}
The normalized MSE of other channels can be similarly calculated as in the above.
Given the total number of subsurfaces $M=40$, we set $M_1=M_2=M/2=20$ for the two distributed IRSs.
Without loss of generality, all the users are assumed to have equal transmit power, i.e., $P_k = P, \forall k$ and the noise power at the BS is set as $\sigma^2_N=-65$ dBm.
%The system operates at a carrier frequency of $6$ GHz with the wavelength of $0.05$ m and the noise power at the BS is set as $\sigma^2_N=-65$ dBm.
Accordingly, the normalized noise power at the BS is given by $\sigma^2=\sigma^2_N/P$.

\begin{figure}
	\centering
	\hspace{-0.1cm}\subfigure[Training overhead versus number of BS antennas $N$.]{
		\begin{minipage}[b]{0.23\textwidth}
			\includegraphics[width=1.05\textwidth]{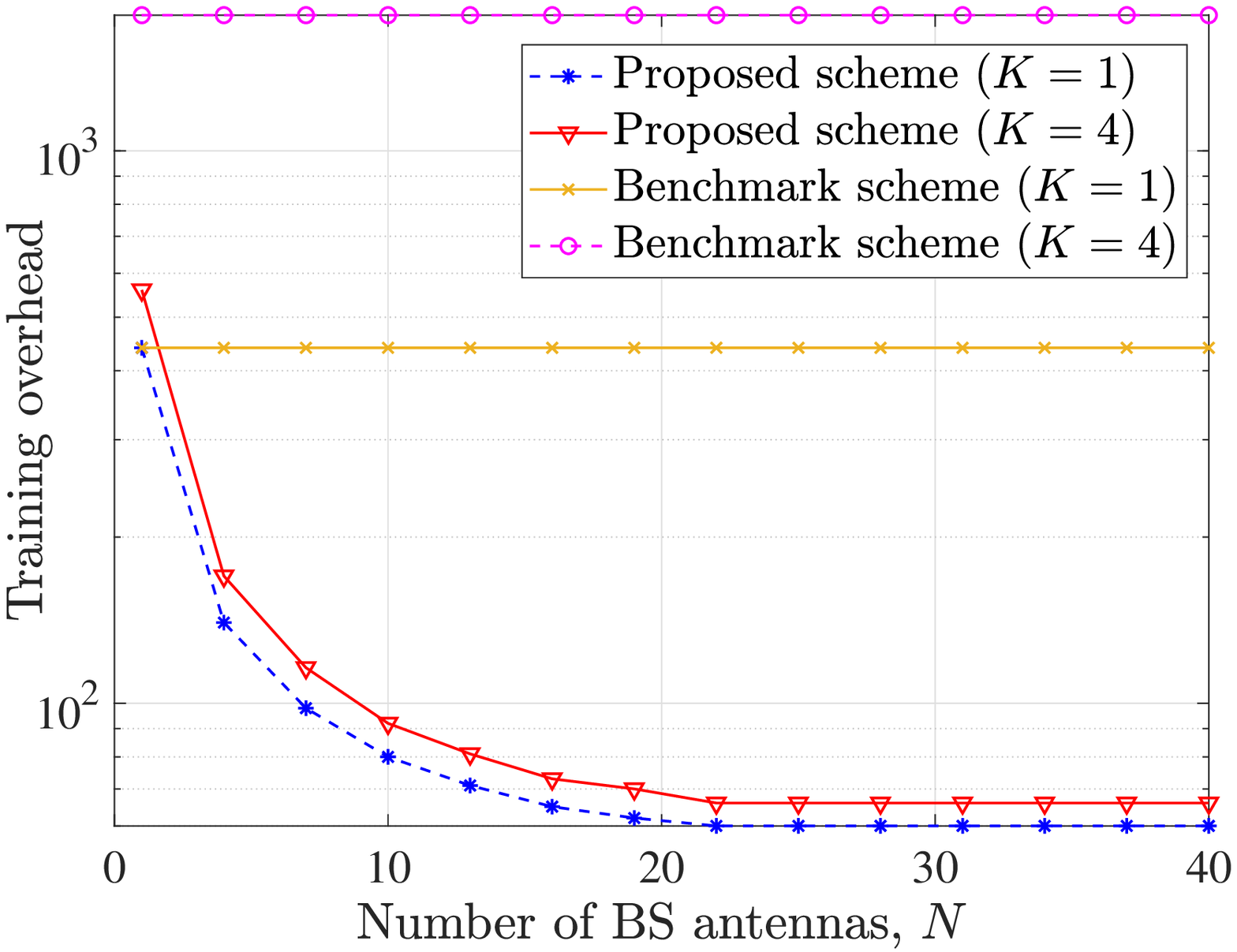}
		\end{minipage}\label{overhead_vsN}
	}%\vspace{0.5cm}
	\hspace{0.1cm}\subfigure[Training overhead versus number of users $K$.]{
		\begin{minipage}[b]{0.23\textwidth}
			\includegraphics[width=1.05\textwidth]{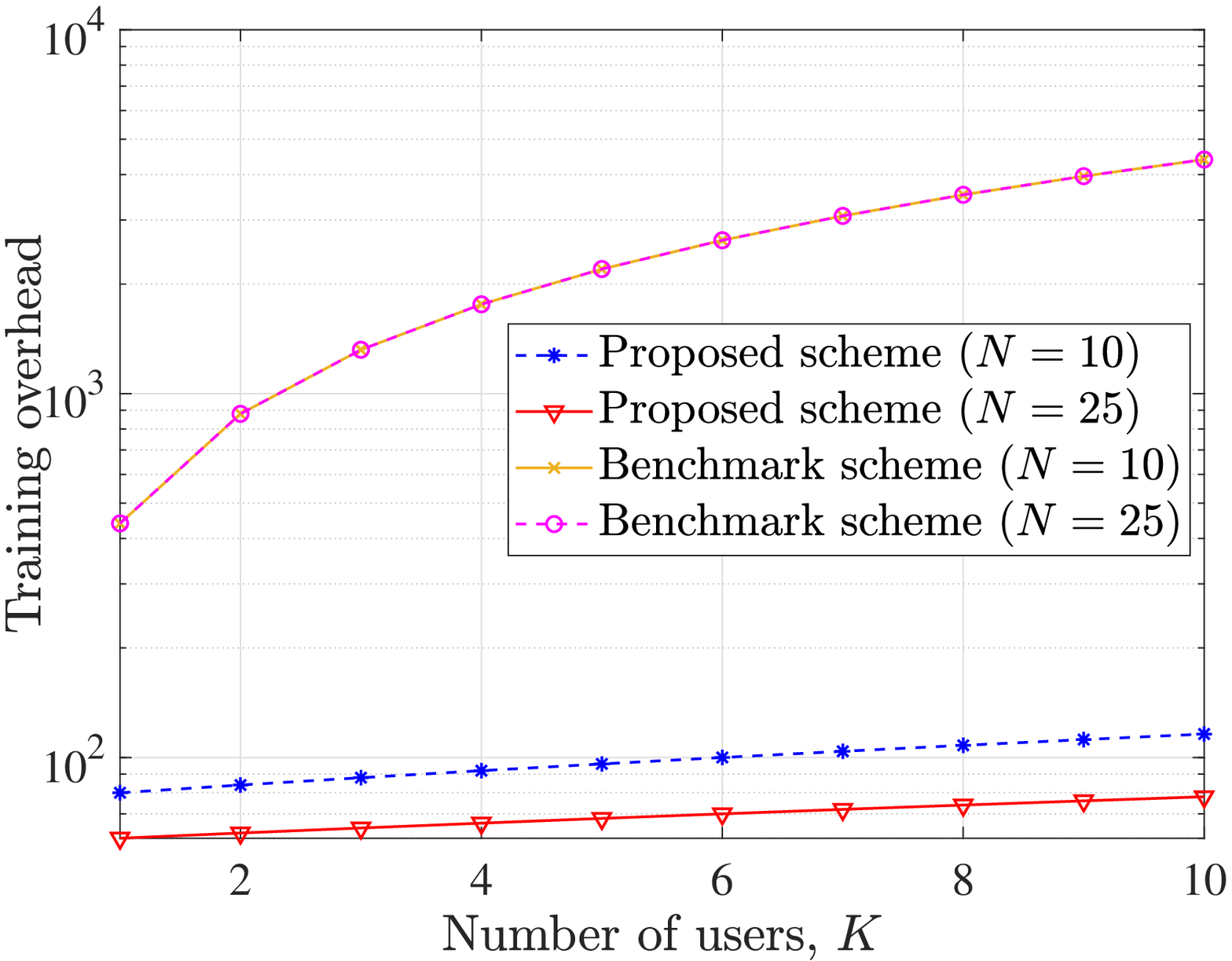}
		\end{minipage}\label{overhead_vsK}
	}
	\setlength{\abovecaptionskip}{-8pt}
	\caption{Training overhead comparison between the proposed scheme and the benchmark scheme based on \cite{you2020wireless}.} \label{overhead}
	\vspace{-0.3cm}
\end{figure}

In Fig.~\ref{overhead_vsN}, we show the required training overhead versus the number of antennas, $N$, at the BS. It is observed that for the proposed channel estimation scheme, the total training overhead decreases with the number of BS antennas $N$,
which is in sharp contrast to the benchmark scheme where its training overhead is independent of $N$.
This is expected since the proposed channel estimation scheme exploits the multiple antennas at the BS with joint IRS channel estimation to reduce training overhead substantially, whereas
in the benchmark scheme the BS estimates its channels associated with different antennas independently in parallel without exploiting the channel relationship between them.
When the number of BS antennas is sufficiently large (i.e., $N\ge \text{max}\{M_1, M_2\}=20$), the minimum training overhead in the proposed scheme reaches its lower bound of $2M_1+M_2+2(K-1)$ pilot symbols.

In Fig.~\ref{overhead_vsK}, we show the required training overhead versus the number of users $K$.
On can observe that the training overhead increment is marginal in the proposed channel estimation scheme as the number of users $K$ increases.
In contrast, the training overhead required by the benchmark scheme increases dramatically with $K$ since it does not exploit the (common) channel relationship among different users.
As such, by fully exploiting the channel relationship between the single- and double-reflection channels as well as among different users,
the proposed scheme achieves much lower training overhead than the benchmark counterpart.

\begin{figure}
	\centering
	\hspace{-0.1cm}\subfigure[Normalzied MSE versus user transmit power $P$ for the single-user case.]{
		\begin{minipage}[b]{0.23\textwidth}
			\includegraphics[width=1.05\textwidth]{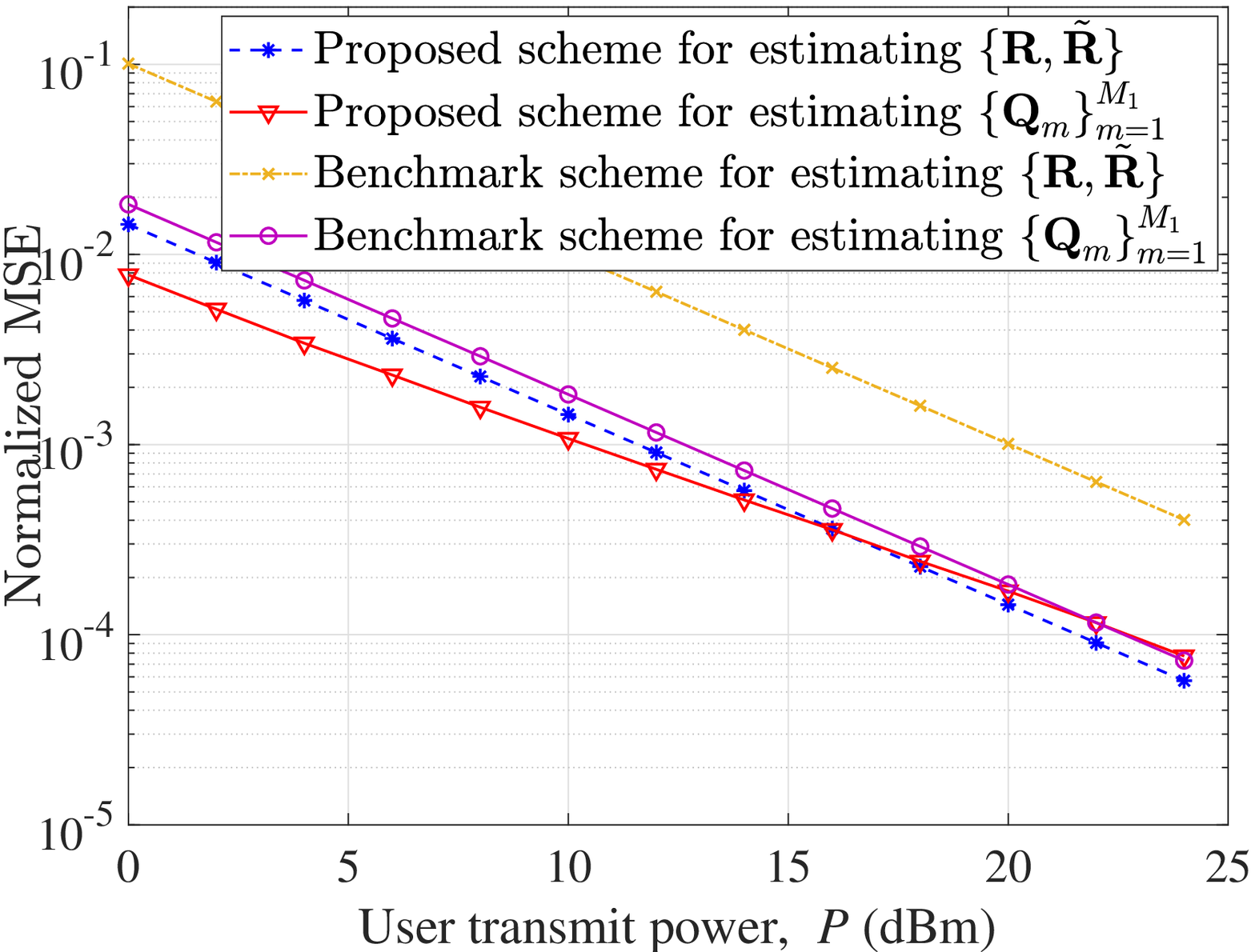}
		\end{minipage}\label{MSE_SNR}
	}%\vspace{0.5cm}
	\hspace{0.05cm}\subfigure[\hspace{-0.05cm}Normalzied MSE versus~user transmit power $P$ for the multi-user case.]{
		\begin{minipage}[b]{0.23\textwidth}
			\includegraphics[width=1.05\textwidth]{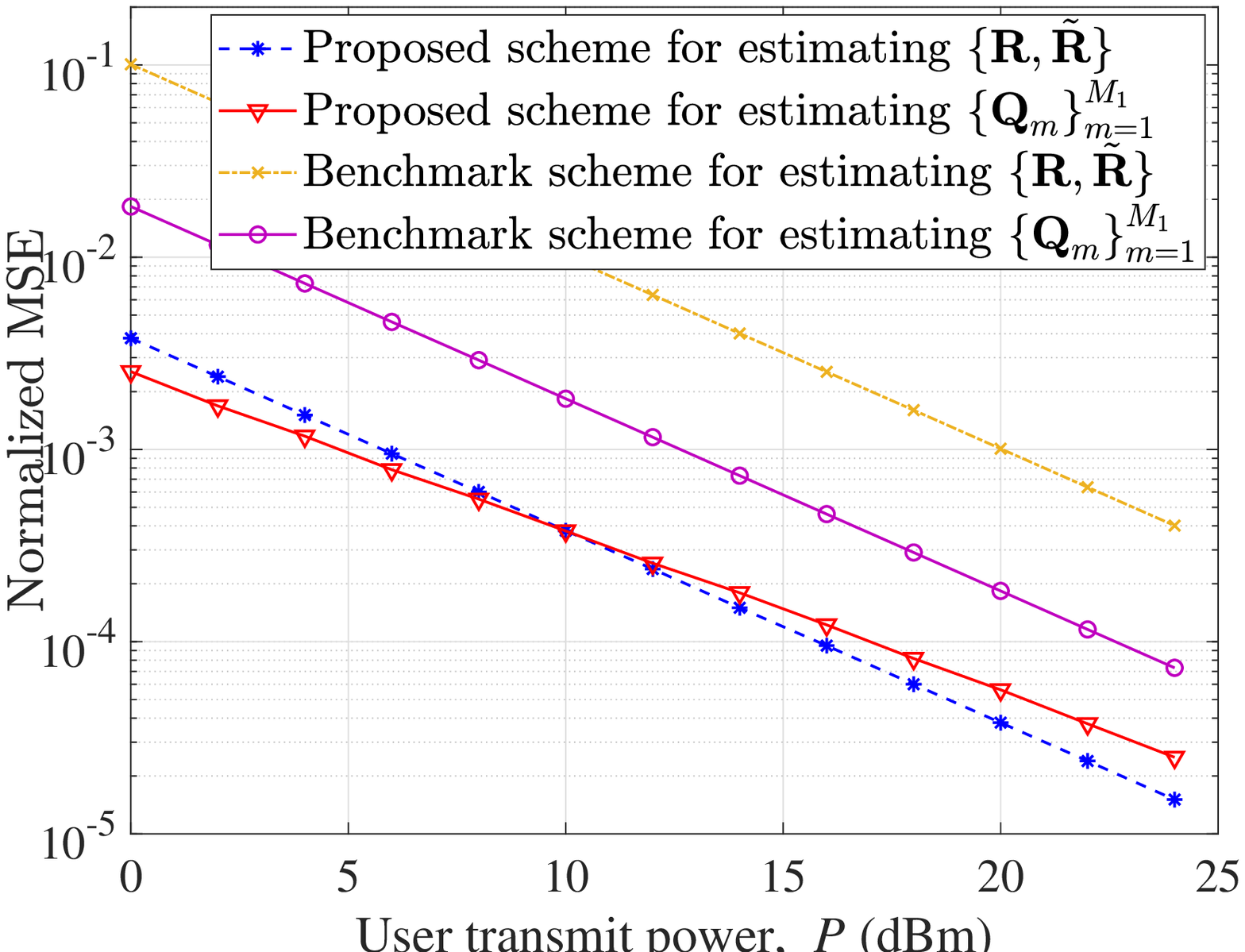}
		\end{minipage}\label{MSE_SNR_MU}
	}
	\setlength{\abovecaptionskip}{-8pt}
	\caption{Performance comparison between the proposed scheme and the benchmark scheme based on \cite{you2020wireless} under the same training overhead.} \label{Performance}
	\vspace{-0.3cm}
\end{figure}

In Figs.~\ref{MSE_SNR} and \ref{MSE_SNR_MU}, we compare the normalized MSE performance of different schemes versus user transmit power $P$ for the single- and multi-user cases, respectively. For fair comparison, we proportionally increase the training overhead of each phase in the proposed channel estimation scheme until reaching the same total training overhead as the benchmark scheme.
It is observed that the proposed scheme achieves much lower MSE than the benchmark scheme, especially for the multi-user case. Moreover, the MSE performance gap between the single- and double-reflection channels in the benchmark scheme is much larger than that in the proposed scheme. This can be understood since the training overhead for the double-reflection channel is of higher order than that for the single-reflection channels in the benchmark scheme ($\frac{1}{4}KM^2$ versus $KM$). In contrast, the proposed scheme achieves balanced MSE performance for the single- and double-reflection channels with the proper proportional training time allocation.

\vspace{-0.2cm}
\section{Conclusions}
In this paper, we proposed an efficient uplink channel estimation scheme for the double-IRS assisted multi-user MIMO system. 
For the single-user case, the higher-dimensional double-reflection channel was efficiently estimated with substantially reduced training overhead by exploiting the property that its cascaded channel coefficients are the scaled versions of those of a lower-dimensional single-reflection channel.
%For the single-user case, the high-dimensional double-reflection channels can be efficiently estimated by exploiting the fact that these cascaded channels are scaled versions of one single-reflection channel; as a result, only the low-dimensional scaling factors need to be estimated to substantially reduce training overhead.
The proposed channel estimation scheme was then extended to the multi-user case by exploiting the fact that the other users' cascaded channels are scaled versions of that of an arbitrary (reference) user's cascaded channel for training overhead reduction.
Simulation results demonstrated the effectiveness of the proposed channel estimation scheme as compared to the existing scheme.

\ifCLASSOPTIONcaptionsoff
  \newpage
\fi

\vspace{-0.2cm}
\bibliographystyle{IEEEtran}
% argument is your BibTeX string definitions and bibliography database(s)
\bibliography{IRS_MIMO}

\end{document}